\documentclass[%
 reprint,
 superscriptaddress,
 amsmath,amssymb,
 aps,
 prl,
]{revtex4-1}

\usepackage{graphicx}
\usepackage{dcolumn}
\usepackage{bm}
\usepackage{xcolor}
\usepackage{ulem}
\usepackage{hyperref}
\usepackage{cleveref}
\usepackage{natbib}

\begin{document}

\title{Nanoscale design of the local density of optical states}

\author{Sandro Mignuzzi}
\email{s.mignuzzi@imperial.ac.uk}

\affiliation{
 The Blackett Laboratory, Department of Physics, Imperial College London, London SW7 2BW, United Kingdom\\
 }

\author{Stefano Vezzoli}
\affiliation{
 The Blackett Laboratory, Department of Physics, Imperial College London, London SW7 2BW, United Kingdom\\
}

\author{Simon A. R. Horsley}
\affiliation{Department of Physics and Astronomy, University of Exeter, Exeter EX4 4QL, United Kingdom}

\author{William L. Barnes}
\affiliation{Department of Physics and Astronomy, University of Exeter, Exeter EX4 4QL, United Kingdom}

\author{Stefan A. Maier}
\affiliation{
 The Blackett Laboratory, Department of Physics, Imperial College London, London SW7 2BW, United Kingdom\\
}
\affiliation{
Chair in Hybrid Nanosystems, Nanoinstitute Munich, Faculty of Physics, Ludwig-Maxilimians-Universit\"at M\"unchen, 80799 M\"unchen, Germany
}

\author{Riccardo Sapienza}
\affiliation{
 The Blackett Laboratory, Department of Physics, Imperial College London, London SW7 2BW, United Kingdom\\
}

\begin{abstract}
We propose a design concept for tailoring the local density of optical states (LDOS) in dielectric nanostructures, based on the phase distribution of the scattered optical fields induced by point-like emitters.
First we demonstrate that the LDOS can be expressed in terms of a coherent summation of constructive and destructive contributions. By using an iterative approach, dielectric nanostructures can be designed to effectively remove the destructive terms. In this way dielectric Mie resonators, featuring low  LDOS for electric dipoles, can be reshaped to enable enhancements of three orders of magnitude. To demonstrate the generality of the method, we also design nanocavities that enhance the radiated power of a circular dipole, a quadrupole and an arbitrary collection of coherent dipoles. Our concept provides a powerful tool for high-performance dielectric resonators, and affords fundamental insights into light-matter coupling at the nanoscale.
\end{abstract}

\maketitle

\section{Introduction}
The radiation properties of an emitter can be controlled by the photonic environment, through the Purcell effect~\cite{Fox2006,Purcell1946,Landi2018,Schmidt2018}.
According to Fermi's golden rule, the decay rate of an emitter is proportional to the local density of optical states (LDOS), which  is the number of electromagnetic modes per unit volume and frequency, at a given point in space. The design of the LDOS finds applications in cavity electro-dynamics~\cite{Gross2018}, lasing~\cite{Pickering2014}, light sources~\cite{Tsakmakidis2016} and solar cells~\cite{Atwater2010}.
Traditional cavities, including laser resonators and photonic crystal cavities, modify the LDOS spectrally, and enhance light emission only on resonance with the cavity mode~\cite{Noda2007}. Due to their large quality factor ($Q$ up to $10^{10}$~\cite{Asano2017}), these microcavities require a stable and precise spectral matching with the emitter.

More recently, nanocavities and nanoantennas have been explored since they sculpt the optical modes down to the near-field~\cite{Koenderink2017, Chikkaraddy2016,Akselrod2014,Sauvan2013}, enabling nanometric optical confinement, without strong spectral bandwidth restrictions~\cite{Koenderink2017,Chikkaraddy2016}.
This confinement can be achieved with metal nanostructures where the sub-wavelength localization of light is achieved through the excitation of surface plasmon-polaritons~\cite{Maier2007,Carminati2015}, and more recently with high-index dielectric nanostructures which confine light via Mie resonances~\cite{Kuznetsov2016}. Dielectrics are advantageous because they feature minimal absorption losses over broad spectral ranges~\cite{Kuznetsov2016}, however, they typically achieve total decay rate enhancements that are two orders of magnitude lower~\cite{Zambrana-Puyalto2015, Krasnok2016, Regmi2016, Cambiasso2017} than their plasmonic counterparts~\cite{Akselrod2014}. Despite several proposals to increase the quality factor of dielectric nanocavities~\cite{Krasnok2016, Rybin2017}, their application to controlling the emission of light is still far from optimized. 

Previously, high decay rates have been achieved in high optical-field regions, the so-called hotspots~\cite{Rutckaia2017}.  In dielectrics, sub-wavelength hotspots have been achieved in nanogaps~\cite{Almeida2004,Robinson2005} where the electric field is boosted, and a decay rate enhancement as high as $\sim 30$ has been demonstrated~\cite{Kolchin2015,Choi2017}. 
Recently, the concept of mode matching between the dipole and the plasmonic antenna was introduced and used to provide some guiding principles to optimize the emission~\cite{Feichtner2017}. 

Here we propose a general and versatile design route for enhancing the LDOS. By means of reciprocity, the LDOS is reformulated in terms of the fields induced by a dipolar emitter in the dielectric environment. We depart from quasi-static approximations, revealing that the phase of the fields plays an important role down to the near-field of an emitter. This allows to construct a computational design method for nanophotonic structures able to enhance the decay rate of arbitrary emitters of about three orders of magnitude. Our framework allows to explain many of the successful nanocavities and nanoantennas geometries used in nanophotonics, and more importantly to design new ones. Although our method is general, here we focus on the specific case of optical nanocavities to highlight the power of our approach to significantly improve existing designs.

\section{Results and discussion}

\begin{figure}[h]
	\centering
		\includegraphics[width=8.3cm, resolution=100]{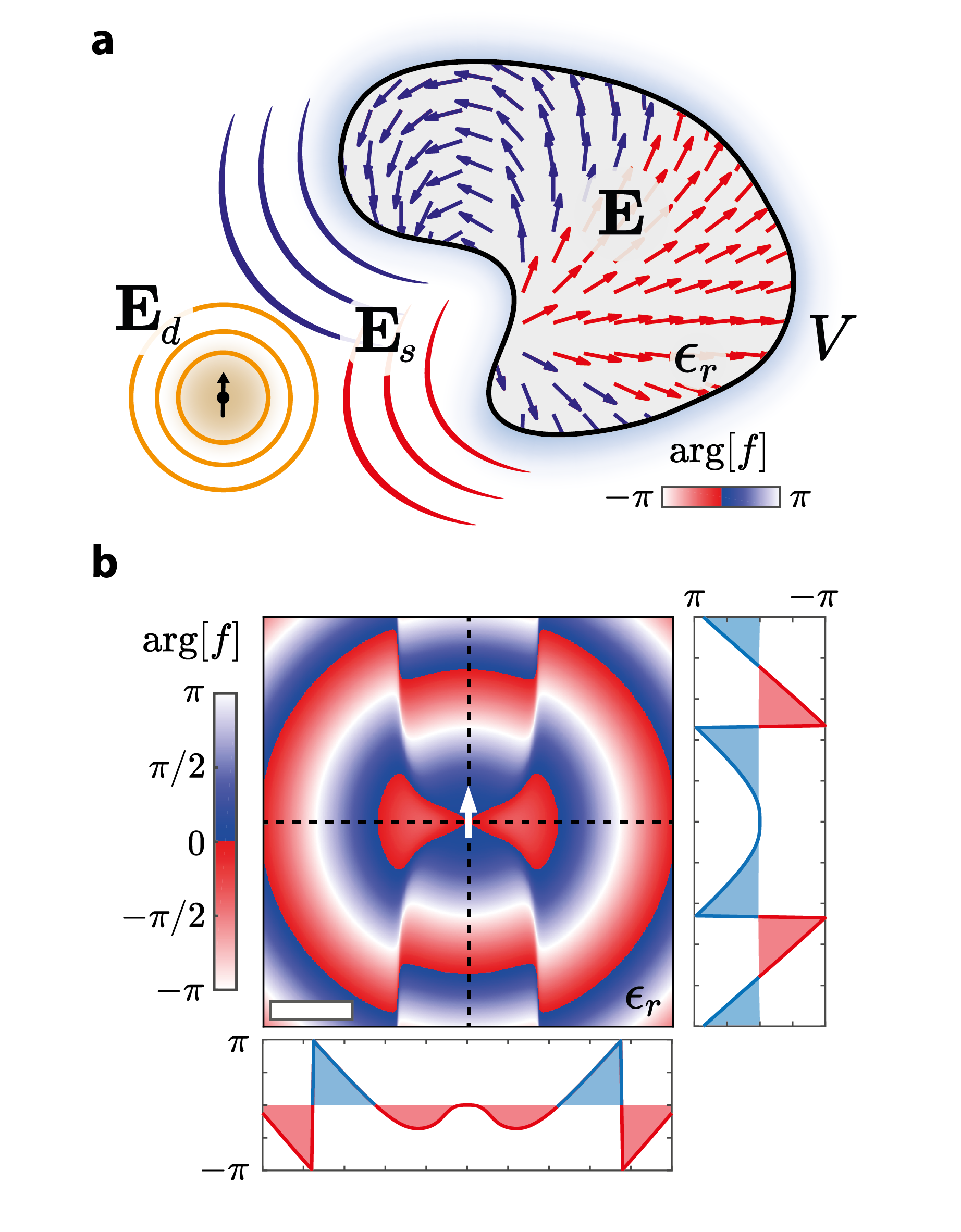}
	\caption{(a) Sketch of an emitter in vacuum in the proximity of a dielectric structure with permittivity $\epsilon_r$. The arrows in the dielectric represent the field components responsible for increasing (blue) or decreasing (red) the decay rate of the emitter. The type of field component is discriminated by ${\rm arg}[f]$ with $f=\textbf{E}_d \cdot \textbf{E}$ defined  in Eq.~\ref{eq:decay2}. (b) A spatial map of ${\rm arg}[f]$ for a dipole emitting at $\lambda_0=800$ nm in a homogeneous medium of $\epsilon_r=4$. The phase is wrapped within $[-\pi, \pi]$.  Scalebar is 200 nm.}
	\label{fig:1}
\end{figure}

We consider the partial (i.e. directionally dependent) LDOS,  $\rho_{\rm p}(\textbf{r}_d)$, for an oscillating electric dipole, at position $\textbf{r}_d$ and along the direction $\textbf{e}_d$, and we compare it to its free space value, $\rho_{\rm 0}$.
The geometry of the problem is sketched in Fig.~\ref{fig:1}a, showing a dipole in vacuum, with dipole moment $\textbf{d}$, in proximity to a dielectric of volume $V$. The dipolar source field 
is back-scattered by the dielectric environment, producing a net scattered field $\textbf{E}_s$ at the site of the dipole. The partial LDOS enhancement, $\rho_{\rm p}/\rho_0$, and the radiative decay rate enhancement, $\gamma/\gamma_0$, depend on this back-action as~\cite{Hecht2006, Carminati2015, Caze2013}:

\begin{equation}
	\frac{\rho_{\rm p}}{\rho_0} = \frac{\gamma}{\gamma_0} = 1+\frac{6 \pi \epsilon_0}{\left|\textbf{d}\right|^2}\frac{1}{k^3} \text{Im}\left\{\textbf{d}^\ast\cdot \textbf{E}_s(\textbf{r}_d)\right\}
	\label{eq:decay}
\end{equation}

\noindent
with $\epsilon_0$ the permittivity in vacuum, $k$ the wavenumber. $\rho_0$ and $\gamma_0$ are the free space LDOS and radiative decay rate respectively.
The sources that produce the scattered fields are the induced currents $\textbf{J}_s$ in the dielectric surrounding the emitter. Such currents are related to the field $\textbf{E}$ inside the dielectric volume by $\textbf{J}_s= -i\omega\epsilon_0(\epsilon_r-1)\mathbf{E}$.
The reciprocity theorem~\cite{Hecht2006} allows Eq. (\ref{eq:decay}) to be re--written as~(see Methods):
\begin{align}
\begin{split}
	\frac{\gamma}{\gamma_0} 
	&= 1+\frac{6 \pi \epsilon_0^2}{\left|\textbf{d}\right|^2}\frac{1}{k^3} ~\text{Im}\left[\int_{V}  d^3\textbf{r}~(\epsilon_r(\textbf{r})-1) ~\textbf{E}_{d^{\ast}}(\textbf{r}) \cdot \textbf{E}(\textbf{r})\right] 
	\label{eq:decay2}
\end{split}
\end{align}
\noindent where the integral extends over the volume $V$ of the dielectric, and $\textbf{E}_{d^{\ast}}$ is the field of the time--reversed dipole moment in free space (i.e. $\textbf{E}_{d^{\ast}}=\omega^2\mu_0\textbf{G}_0(\textbf{r},\textbf{r}_d)\cdot\textbf{d}^{\ast}$, with $\textbf{G}_0$ the free space Green function).  
The derivation is further detailed in the Supplementary Information, where an equivalent Green function formalism approach is also described.
In the following we will assume the environment is composed by dielectrics with $\epsilon_r$ real, i.e. without ohmic losses. The total decay rate considered is therefore only radiative.
As illustrated in Fig.~\ref{fig:1}a, the decay rate is determined by the coherent contributions of spatially distinct field elements in the dielectric medium. Each field element is responsible for driving or damping the emitter (blue and red respectively in Fig.~\ref{fig:1}a), depending on the sign and amplitude of ${\rm Im}[\textbf{E}_{d^{\ast}} \cdot \textbf{E}] = {\rm Im}[f]$, where we define $f = \textbf{E}_{d^{\ast}} \cdot \textbf{E}$.  Here we are assuming $\epsilon_r$ real. The amplitude of $f$, i.e. $|f|$, can be controlled by matching the (vacuum) field distribution of the emitter $|\textbf{E}_{d^{\ast}}|$ with that of the induced fields $|\textbf{E}|$, as shown in Ref.~\cite{Feichtner2017}. On the other hand, the phase of $f$, i.e. ${\rm arg}[f]$, determines the sign of {\rm Im}[f], and thus dictates the enhancement or suppression of the radiative decay. Regions of dielectric where ${\rm arg}[f]\in [0,\pi]$ (i.e. ${\rm Im}[f]>0$) (shown blue) give a positive contribution, whereas regions where ${\rm arg}[f]\in [-\pi,0]$ (i.e. ${\rm Im}[f]<0$) (shown red) give a negative contribution, and therefore should be removed if the decay rate is to be enhanced. 

\begin{figure}
	\centering
		\includegraphics[width=8.3cm, resolution=100]{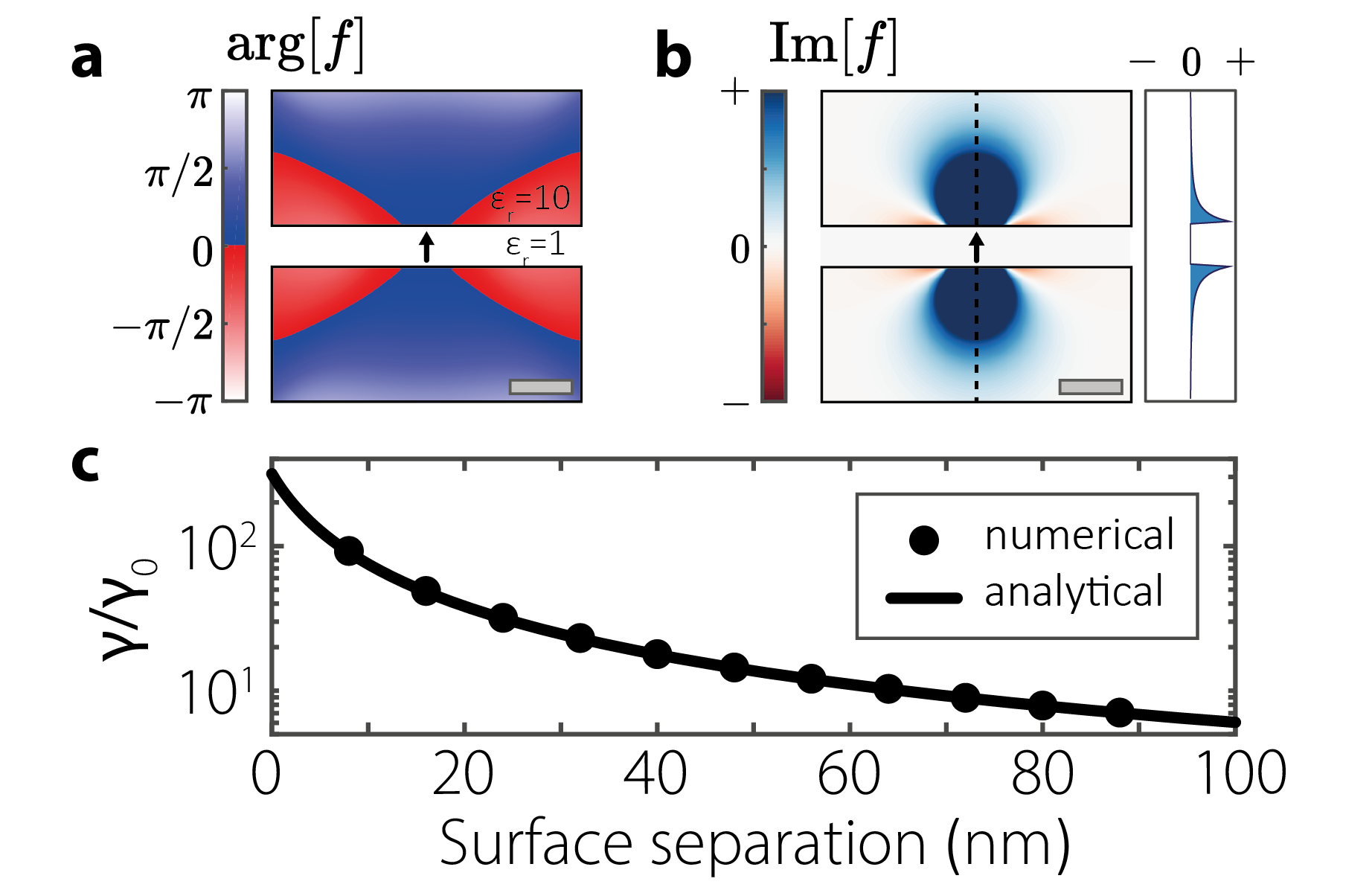}
	\caption{LDOS between two infinite parallel plates of $\epsilon_r=10$, for a perpendicular dipole ($\lambda_0=800$~nm) in the middle of the air gap. 
    (a) The spatial distribution of ${\rm arg}[f]$, 
    (b) of ${\rm Im}[f]$ and the relative cross-section along the dotted line, and  
    (c) decay rate enhancement as a function of the separation between the plates calculated both analytically (line)~(see Supplementary Information) and solving Eq.\ref{eq:decay2} numerically. Scale bar is 50 nm.}
	\label{fig:2}
\end{figure}

In Fig. \ref{fig:1}b we show ${\rm arg}[f]$ for a dipole in a homogeneous medium, calculated analytically~(see Supplementary Information). The phase of the fields is relevant even within the near-field region of the dipole, where one would assume optical retardation is negligible.  As illustrated by the cross sections, in the near-field (within a distance $\sim\lambda_0/2{\rm n}$, where n is the refractive index, which is the size of the scale bar in Fig. 1b) ${\rm arg}[f]$ is positive along the longitudinal direction, but negative along the transversal direction. 
The complex function $f$ is key to the nanoscale design of structures for LDOS control. 
For LDOS enhancement, the damping contributions (red regions in Fig. \ref{fig:1}b) to the integral in Eq. \ref{eq:decay2} could be removed by physically removing dielectric material in the regions where ${\rm arg}[f]\in [-\pi,0]$. The fields and the function $f$ can then be re-calculated and the process repeated iteratively so as to optimize the degree of enhancement.

We first test this concept on a simple vacuum nanogap surrounded by two dielectric half-spaces with a high refractive index~\cite{Robinson2005,Jun2009}, as illustrated in Fig. \ref{fig:2}, which can be solved analytically~(see Supplementary Information).
We consider a dipole source (wavelength $\lambda_0$ = 800~nm) located at mid-gap, and oriented  perpendicular to the interfaces. The permittivity is $\epsilon_r=1$ at the dipole position and $\epsilon_r=10$ within the dielectric plates. 
The spatial distributions of ${\rm arg}[f]$ and ${\rm Im}[f]$, for a gap width of 32~nm, are shown in Fig. \ref{fig:2}a and b respectively. 
As seen in Fig. \ref{fig:2}b, the sign of ${\rm Im}[f]$ is strongly correlated with the distribution of the phase (Fig. \ref{fig:2}a), whereas the amplitude of ${\rm Im}[f]$ becomes stronger in the proximity of the dipole, where the field is highest.
We numerically calculate $f(\textbf{r})$ and the decay rate, as prescribed by Eq.\ref{eq:decay2}, using FDTD.
The decay rates calculated numerically are in very good agreement with those calculated analytically~(see Supplementary Information), as shown in Fig.~\ref{fig:2}c. 
As shown in Fig. \ref{fig:2}b, ${\rm Im}[f]$ is mostly positive (blue), leading to a strong enhancement of the decay rate:
the nanogap is effectively removing regions of dielectric where polarization currents would give a negative contribution to the LDOS (red).

\begin{figure}
	\centering
		\includegraphics[width=8.3cm, resolution=100]{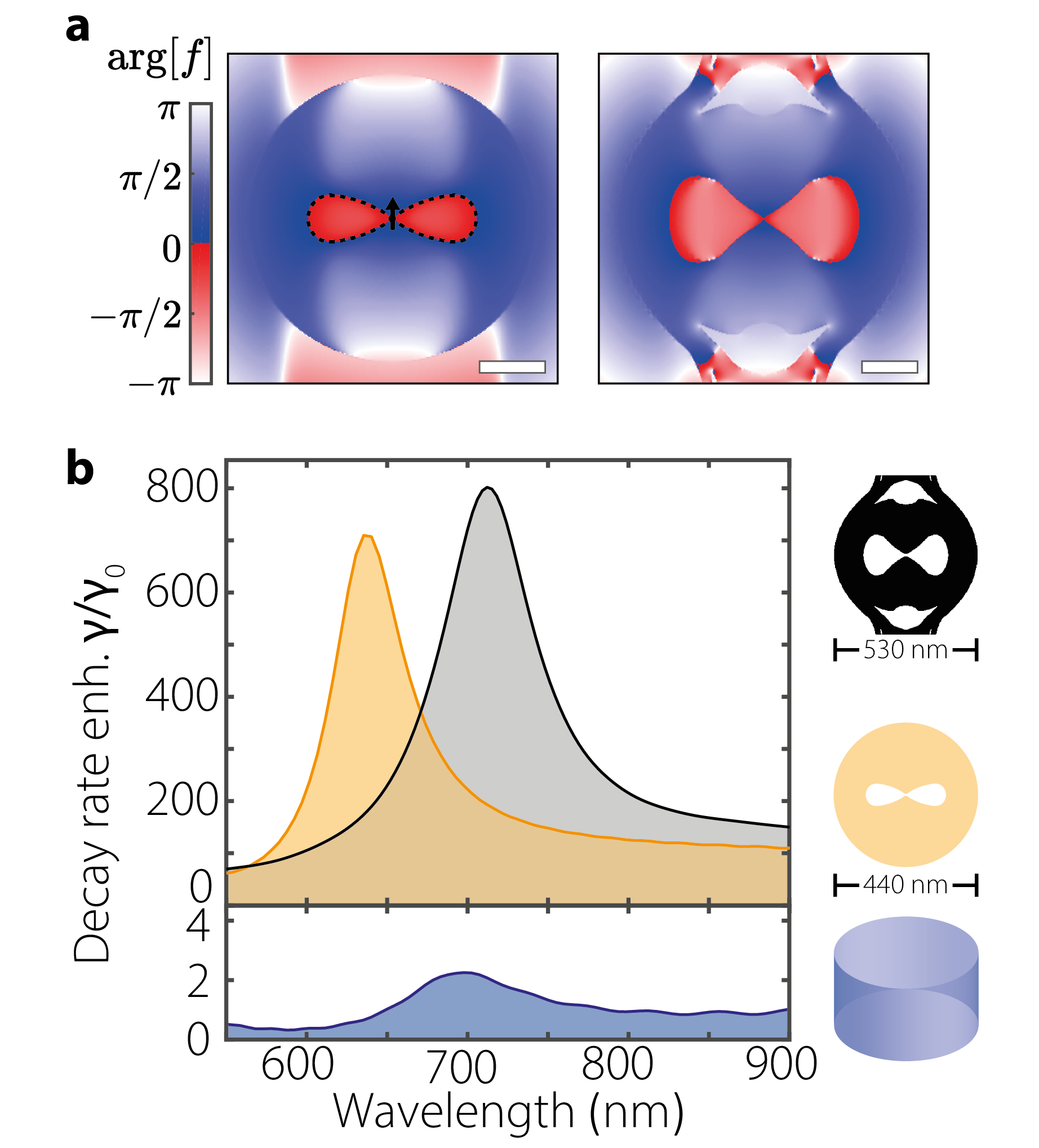}
	\caption{(a) On the left: phase map ($\arg(f)$) of a GaP nanodisk resonator (440 nm diameter, 50 nm height), when excited by a dipole ($\lambda = 700$~nm) in its center at mid-height, surrounded by vacuum. The dotted line indicates the out-of-phase components that are removed to implement the phase design. Scale bar is 100~nm.  On the right: phase map of a resonator designed by an iterative procedure, featuring a double-hole structure.  (b) Decay rate enhancement for a full nanodisk resonator (blue), a  perforated one (yellow) and a resonator designed by an iterative procedure (black). }
	\label{fig:3}
\end{figure}

We now discuss the design of nanoresonators, compatible with current nanofabrication technology, for the efficient enhancement of the decay rate. 
We consider a simple resonator geometry, a high index dielectric nanodisk with a Mie resonance, which has been recently demonstrated to effectively confine radiation \cite{Miroshnichenko2015,Yang2018}.
Such sub-wavelength structures typically have low quality factors (Q$<30$) \cite{Rybin2017}, confine the field to a volume of $\sim \lambda^3$, and consequently feature poor decay rate enhancements. In Fig. \ref{fig:3}a we illustrate a spatial map of ${\rm arg}[f]$ for a gallium phosphide (GaP) dielectric nanodisk, excited at its anapole resonance~\cite{Miroshnichenko2015,Yang2018} by a dipole at the center of the disk, at half its height. 
The anapole resonance arises, at certain wavelengths given by the particle size, from a superposition of a dipolar and toroidal mode, whose far field components cancel out. In other terms, the energy is effectively stored within the particle.
The destructive contributions (red) to the LDOS in Fig.~\ref{fig:3}a, prevents large decay rate enhancements. As shown in of Fig.~\ref{fig:3}b (blue curve) the decay rate enhancement is peaked at $\sim$ 700 nm, and is never larger than  $\sim 2$ times. 
When the dielectric material with negative ${\rm arg}[f]$, i.e. the red double-hole region between the dotted lines in Fig.~\ref{fig:3}a, is removed, 
the decay rate increases to $\gamma / \gamma_0 = 708$, as illustrated in Fig.\ref{fig:3}b (yellow curve).
The blue-shift of the resonance is due to a lower effective refractive index.
In our design the double-hole is bridged via a nanogap of 8 nm, which is feasible with He focused ion beam drilling~\cite{Fang2015}.

This nanocavity design, based on a visual inspection of ${\rm arg}[f]$, perturbs the modes of the structure so that the distribution of ${\rm arg}[f]$ is no longer as it was in Fig.~\ref{fig:3}a before the introduction of the holes. 
The design can be improved further through an iterative procedure where for each step the permittivity $\epsilon$ is gradually changed by an infinitesimal amount $\pm\delta\epsilon$, depending on whether $\mathrm{Im}[f]$ is positive or negative; all points in the dielectric volume surrounding the dipole emitter are considered in this way.
By using a Born approximation it can be demonstrated~(see Methods) that the change in the integral of Eq.~\ref{eq:decay2} is proportional to  
$~\delta\epsilon~{\rm Im}\left[\textbf{E}_{\ast}(\textbf{r})\cdot\textbf{E}(\textbf{r})\right]$, where $\textbf{E}_{\ast}$ is the total electric field due to the time reversed dipole moment $\textbf{d}^{\ast}$.
The electric field is numerically recalculated after each change to the permittivity profile, so that the LDOS is optimized at $\textbf{r}_{d}$, and the final distribution of currents serves to enhance the emission (${\rm arg}[f]>0$) as much as possible.
The graded index resulting from the optimization is then converted to a binary index, for air ($\epsilon_r=1$) and GaP ($\epsilon_r\sim10$), using a threshold of $\epsilon_r=5.5$, so as to enable implementation with a homogeneous dielectric. In particular, the structure is planarized using the index profile at mid-height of the slab. The graded index is then converted to a binary index by using the index threshold value found by FDTD by maximizing the value of the LDOS.
By using this iterative method, a GaP slab is transformed into a closed ring-like structure, shown in black in the inset of Fig. \ref{fig:3}b. This is similar to the nanodisk with the eight-shaped holes that we discussed previously (yellow), but with a further boost of $\gamma / \gamma_0$ to $\sim 800$. Interestingly, this shape is strongly reminiscent of the phase distribution already encountered in Fig. \ref{fig:1}b, which re-enforces our interpretation of the optimization procedure as a \textit{phase design}. Note that in both cases the mode of the antenna is weakly perturbed and retains a Q of $\sim 10$. 
The decay rate could be further increased with a double-tip with a smaller gap or by increasing the index contrast.
The structures shown in Fig.~\ref{fig:3}b carry a strong resemblance to structures already deployed in nanophotonics, such as bow-tie antennas~\cite{Choi2017}, double hole apertures~\cite{Regmi2015,Gondarenko2008}, dielectric slots~\cite{Robinson2005} and pointed probes used in nanospectroscopy~\cite{Sanz-Paz2018, Mignuzzi2017,Berthelot2014}. The experimental success of this range of nanostructures can thus be re-interpreted in the light of our model as a result of the phase distribution in the medium surrounding the dipole. Interestingly, our approach also rationalizes the designs resulting from computationally expensive methods, such as genetic algorithms~\cite{Gondarenko2006} and inverse designs based on iterative methods~\cite{Lu2011}.

\begin{figure}[!t]
	\centering
		\includegraphics[width=8.6cm, resolution=100]{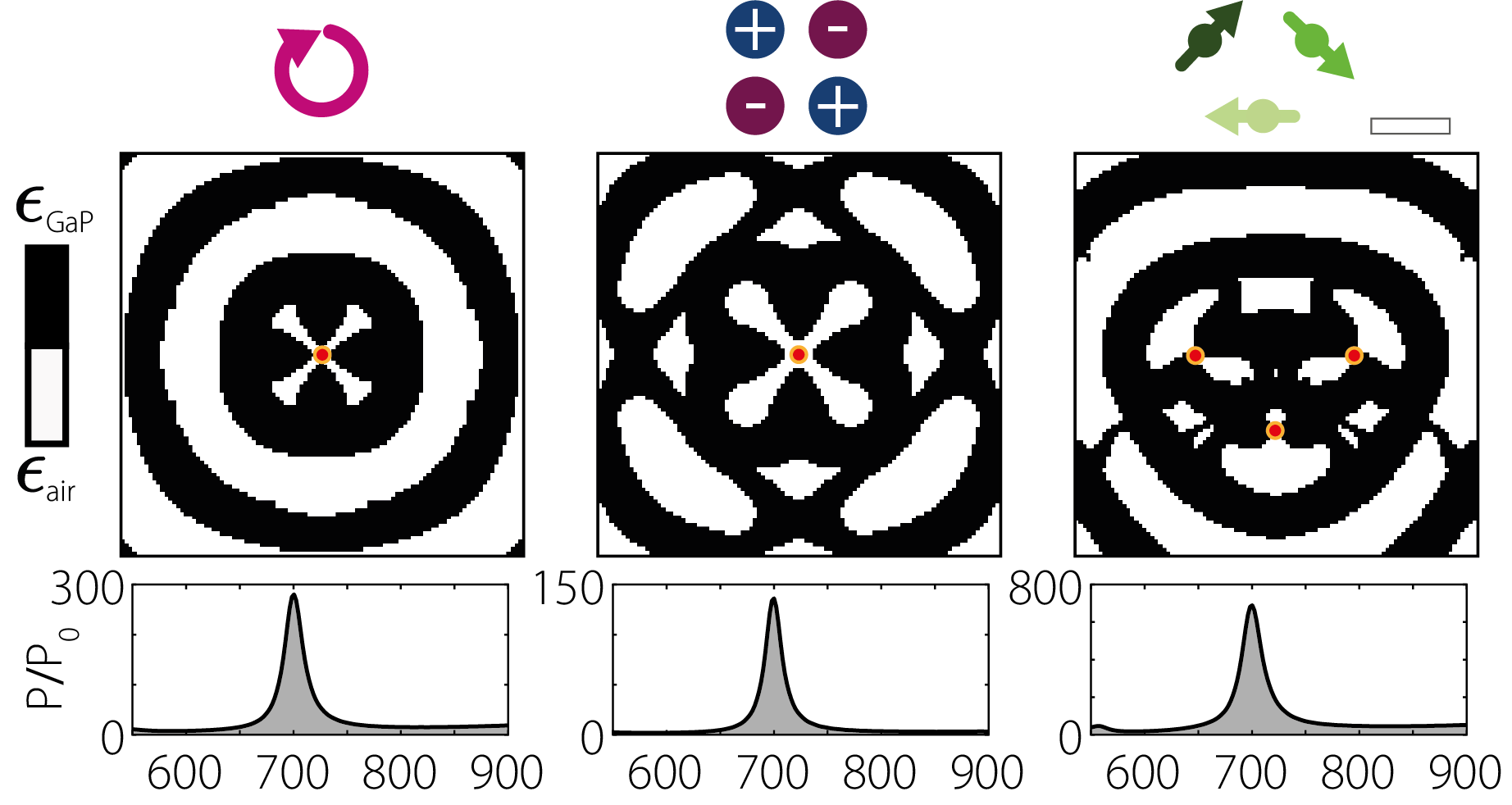}
	\caption{Design of nanocavities for different sources. From left to right: circular dipole, quadrupole, three coherent dipoles (at a distance of 400 nm from each other). The nanoresonators are designed starting from a 50~nm GaP slab, and optimized for a wavelength of 700~nm. Scale bar is 200 nm. The radiated power enhancement ${\rm P/P_0}$ is shown in the bottom panels.}
	\label{fig:4}
\end{figure}

Finally, we illustrate the generality and versatility of our design approach by considering a few non trivial examples beyond the simple case of a single dipole emitter.  Fig.~\ref{fig:4} illustrates the results of the optimization procedure for a circular dipole (a), a quadrupole point-like emitter (b) and a collection of three coherent dipole emitters located at the vertices of a triangle (c). The nanocavity design is obtained, as before, by gradually changing the dielectric constant of an infinite 50 nm thick slab of GaP, within a region of $1 \times 1~\rm{\mu m^2}$, in order to reduce the out-of-phase components (the threshold used for the binary index structure is $\epsilon_r=5.5$). 
The Purcell enhancement obtained for the final structures is illustrated in the lower panel of Fig.~\ref{fig:4}: it reaches a maximum of 280 for the circular dipole and of 135 for the quadrupole, whereas for the three dipole system the radiated power increases by 680 times.

In conclusion, we described a design route for enhancing the LDOS and the radiative decay rate in dielectric nanoresonators based on the spatial distribution of the phase of the induced polarization currents. This is intuitive, versatile and computationally efficient. We obtain decay rate enhancements of 800 for an electric dipole in an improved Mie resonator and similar values for more complex architectures, i.e. circular dipoles, quadrupoles and ensembles of coherent emitters.
The method can be extended to lossy dielectrics, plasmonic materials, and is valid also for larger structures, as for example nanoantennas embedded in micro-resonators~\cite{Hu2017a}.
In particular, the coherent design can improve nanocavities for weakly emitting point sources, such as magnetic and multipole molecular transitions, or for acoustics \cite{Landi2018} and elastic waves \cite{Schmidt2018}. Extreme LDOS enhancements have important applications for single-molecule spectroscopy and sensing, and open a path towards room temperature quantum coherence for nanoscale quantum optics, strong coupling and single photon non-linearities.

\section{Acknowledgement}

The authors acknowledge V. Giannini and J. Cambiasso for fruitful discussion. S.M., S.V., S.A.M., and R.S. acknowledge funding by EPSRC (EP/P033369 and EP/M013812). S.A.M acknowledges the Lee-Lucas Chair in Physics. W.L.B. would like to acknowledge support through an ERC-funded project Photmat (ERC-2016-AdG-742222).  S.A.R.H. acknowledges funding from the Royal Society and TATA.  W.L.B. and S.A.R.H. are indebted to Willem Vos for his invaluable insights and inspiration in thinking about LDOS concepts.

\section{Supporting Information}

Details on the derivation of Eq.2, analytical derivation of ${\rm arg}[f]$ for a homogeneous medium and a dielectric nanogap, and the numerical optimization procedure based on the Born approximation.

\bibliographystyle{apsrev4-1} 
\bibliography{biblio}

\end{document}